\documentstyle[preprint,tighten,floats,prd,aps]{revtex}
\input epsf

\newcommand{\be}{\begin{eqnarray}}
\newcommand{\ee}{\end{eqnarray}}
\begin{document}

\draft

\preprint{arch-ive/9806443 \hspace{104mm} UAHEP984}

\title{
Gluino Contribution to the 3-loop Quark Mass Anomalous Dimension
 in the Minimal Supersymmetric Standard Model}
\author{L.\ Clavelli and L.\ R.\ Surguladze}
\address{Department of Physics \& Astronomy, University of Alabama,
                   Tuscaloosa, AL 35487, USA}
\date{June 7, 1998}
\maketitle
\begin{abstract}
We deduce the gluino contribution to the three-loop QCD quark mass anomalous
dimension function within the minimal supersymmetric Standard Model (MSSM)
from its standard QCD expression. This work is a continuation of the program
of computation of MSSM renormalization group functions.
\end{abstract}


The renormalization group method \cite{RG} is a powerful tool for the study
of many physically interesting quantities within the Standard Model
and beyond. Although experimental measurements at the highest available
energy are consistent with the standard model \cite{exp}, the observed
relationship of the strong coupling constant at the Z and the weak
angle as well as the value of the $m_b/m_{\tau}$ ratio vis-a-vis the
top quark mass remain strong indications of a supersymmetric (SUSY)
grand unification above $10^{16} GeV$ and a SUSY threshold for squarks
and sleptons in the 0.1 to 1 TeV region. To use the renormalization group
method to study the above and other quantities one needs to know
the renormalization group functions - $\beta$ - function and quark
mass anomalous dimension $\gamma_m$. In our previous work we have 
calculated the three-loop QCD $\beta$ function with gluino contribution
included \cite{betagluino}. In the present work we deduce the 
three-loop quark mass anomalous dimension function including the
gluino contribution.
Here we use the known result for the standard QCD expression
of the three-loop quark mass anomalous dimension \cite{tarasov}.
The two-loop contributions in the ratio $m_b/m_{\tau}$ turned out 
to be 20\% of the leading contribution \cite{Nanop}. For this reason a
full calculation of the three-loop contribution would be useful.

The quark mass anomalous dimension is defined as usual
\be
    -\mu^2 \frac{\partial\ln m}{\partial \mu^2} =  \gamma_m(\alpha_s),
\ee

The renormalization of a quark mass within the MS type framework 
\cite{drg,MSB} has
the following form

\be 
m^{B} = Z_m m = m \left[ 1+ \sum_{i}^{\infty} \frac{a_{i}(\alpha_s)}
                                               {\varepsilon^{i}}\right]
\ee
where B indicates the ``bare'' mass. 
The anomalous dimension function is determined by the lowest order pole
term in the quark mass renormalization constant within the MS framework.
That is, 

\be
\gamma_m(\alpha_s) = \alpha_s 
         \frac{\partial a_{1}(\alpha_s)}{\partial \alpha_s}
       =\gamma_1 \frac{\alpha_s}{4\pi}
           +\gamma_2 (\frac{\alpha_s}{4\pi})^2
                +\gamma_3 (\frac{\alpha_s}{4\pi})^3- \ldots
\ee  

The $a_i$ coefficients are related via a renromalization group equation
that serves as a powerful check of the calculation.

\be
\biggl( \gamma_m(\alpha_s) - \beta (\alpha_s) \frac{\partial}
              {\partial \alpha_s} \biggr)
        a_{i}( \alpha_s )=-\alpha_s \frac{\partial}{\partial \alpha_s }
           a_{i+1}( \alpha_s ),
\ee 
where the three-loop QCD $\beta$ function including the gluino contribution
and ignoring squarks is defined as follows \cite{betagluino}. 
\be
    \mu^2\frac{\partial\alpha_s}{\partial \mu^2} = \alpha_s \beta(\alpha_s),
\ee
where
\be
    \beta(\alpha_s) =   \beta_1 \frac{\alpha_s}{4\pi} 
                     + \beta_2 \biggl(\frac{\alpha_s}{4\pi}\biggr)^2
                    + \beta_3 \biggl(\frac{\alpha_s}{4\pi}\biggr)^3
                         + \ldots   .
\ee
with
\be
 \beta_1 = -{11 \over 3} C_A + {4 \over 3}
  \biggl(N_f T + \frac{n_{\tilde{g}}}{2}C_A\biggr),
\ee
\be
 \beta_2 = - \frac{34}{3}C_A^2
   +\frac{20}{3}\biggl(N_f TC_A + \frac{n_{\tilde{g}}}{2}C_A^2\biggr)   
   +4\biggl(N_f TC_F + \frac{n_{\tilde{g}}}{2}C_A^2\biggr),  
\ee
\begin{eqnarray}
\lefteqn{\hspace{-19mm}
   \beta_3 = - {2857 \over 54} C_A^3 - N_fT(2C_F^2
         - \frac{205}{9} C_F C_A - \frac{1415}{27} C_A^2)
             -(N_fT)^2 (\frac{44}{9} C_F+\frac{158}{27} C_A)}
                                                     \nonumber\\
 && \quad \hspace{15mm}
    + {988 \over 27}n_{\tilde{g}} C_A^3
     - n_{\tilde{g}} N_f T ({224 \over 27}C_A^2 +{22 \over 9} C_A C_F)
      - {145 \over 54}n_{\tilde{g}}^2 C_A^3.               
\label{finalbeta} 
\end{eqnarray}

The MS renormalization of quark mass can be expressed as the following multiplicative renormalization 

\be
Z_m = Z_{\overline{\psi}\psi} \times Z_{2}^{-1}
\ee

Where $Z_2$ is the quark propagator renormalization constant and
$Z_{\overline{\psi}\psi}$ renormalizes the quark propagator with
$\int \overline{\psi}(x)\psi(x) dx$ operator insertion.
\begin{figure}
\hskip 3.0cm
\epsfxsize=3.5in \epsfysize=2in \epsfbox{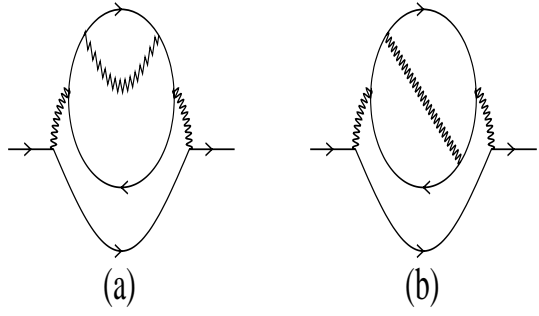}
\caption{Three-loop graphs giving nontrivial contribution to the quark
mass anomalous dimension function with the gluino included. Wavy lines denote
gluons and the solid loop corresponds to quark or gluino.}
\label{Fig.1}
\end{figure}
The gluino contributions to the above renormalization constants are in 
one-to-one correspondence with quark loop graphs and differ from them only 
by color and symmetry factors. Our procedure, as used to determine the 
gluino contributions to Z decay at four-loop level \cite{Zgluino} and
 to the $\beta$-function at three-loop level, is to decompose the known
 QCD results into contributions from separate graphs. Then one can determine
 the color factors that relate each graph with an internal quark loop to
 the corresponding graph with a gluino loop.
For instance, a subgraph consisting a simple quark loop has the color
 factor
\begin{displaymath}
N_f \mbox{Tr}(T^aT^b)=N_f T\delta^{ab}
\end{displaymath}
while the color factor for a simple gluino loop has the color
 (and symmetry)
factor
\begin{displaymath}
\frac{n_{\tilde{g}}}{2} \mbox{Tr} (F^aF^b) = \frac{n_{\tilde{g}}}{2}
                                                            C_A \delta^{ab}
\end{displaymath}
The relative factor 1/2 is due to the Majorana nature of the gluino. 
Here $T^a$ and $F^a$ are the gauge group generators in the fundamental
 and adjoint representations respectively. For the gauge group SU(N)
 they satisfy
\begin{displaymath}
\mbox{Tr}(T^aT^b)=T\delta^{ab}
\end{displaymath}
\begin{displaymath}
\mbox{Tr} (F^aF^b) = N\delta^{ab}
\end{displaymath}
Thus for
graphs with a simple fermion loop subgraph one obtains the gluino 
contribution
by making the substitution
\begin{displaymath}
N_f T \rightarrow N_f T + \frac{n_{\tilde{g}}}{2} C_A
\end{displaymath}
Here $n_{\tilde{g}}=0$ is the Standard Model limit and $n_{\tilde{g}}=1$
corresponds to the minimal SUSY extension with one octet of gluinos.

The full set of three-loop graphs contributing in the standard QCD
$Z_2$ and $Z_{\overline{\psi}\psi}$ were given in ref. \cite{tarasov}.
We reanalyse all of the graphs up to and including three-loop level
and added graphs with gluino propagators. The calculation of SU(N)
group factors revealed that all but two graphs with the gluino contribution
can easily be obtained from the known results by simply replacing
$N_f$ to $N_f +n_{\tilde{g}}C_A$. However, for the graph of Fig.1a
we found that
in order to take the particular topology into account one needs to make the 
following replacement in the standard QCD result for this graph
\be
N_f \rightarrow N_f + \frac {C_{A}^2}{2TC_{F}}=N_f + \frac {27}{4}
\ee
The replacement for the graph in Fig.1b looks like
\be
N_f \rightarrow N_f + \frac {C_{A}^2}{4T(C_{F}-C_{A}/2)}=N_f - 27.
\ee

If the gluino lies above the squark, the current calculation provides the
 contribution from gluinos alone up to three-loop order. This is a gauge
 invariant
subset of the full SUSY three-loop graphs and leaves a vastly reduced number of
 graphs (those with one or more internal squark lines) still to be calculated 
at this order. If the gluino lies lower in mass than the squarks, the current
 calculation produces the full SUSY anomalous dimension of quark mass up to
and including the three loop order in the region up to the squark mass scale.

We obtain the following result for three-loop quark mass anomalous dimension
function with gluino included.

\be
\gamma_1 = 3C_F
\ee

\be
\gamma_2 = C_F \biggl[ \frac{3}{2}C_F+\frac{97}{6}C_A-\frac{10}{3}
                 \biggl( TN_f + \frac{n_{\tilde{g}}}{2}C_A\biggr) \biggr]
\ee

\begin{eqnarray}
\lefteqn{\hspace{-27mm}
     \gamma_3 = C_F \biggl[ 
        \frac{129}{2} C_F^2 - \frac{129}{4} C_F C_A
        -C_F T N_f \biggl( 46 +48 \zeta(3) (C_A-C_F) \biggr)                                                                +\frac{11413}{108}C_{A}^2 - \frac{556}{27} C_A T N_f }
                                                    \nonumber\\
  && \quad
        -\frac{140}{27} \biggl(T N_f+\frac{n_{\tilde{g}}}{2}C_A \biggr)^2 
                                              \biggr]
       -n_{\tilde{g}} C_F
        \biggl[ \frac{1}{2} C_F C_A + \frac{1771}{54} C_A^2 \biggr]
\label{finalgamma} 
\end{eqnarray}

The eigenvalues of the Casimir operators for the adjoint ($N_A=8$) and the
fundamental ($N_F=3$) representations of SU$_{\mbox{\scriptsize c}}$(3) are
\begin{equation}
C_{A}=3,\hspace{2mm}C_{F}=4/3, \hspace{2mm} \mbox{and} \hspace{2mm}
 T=1/2.
\label{eq:casimirsnum}
\end{equation}
We obtain the following values for the above perturbative coefficients
of the $\gamma$-function.

\be
\gamma_1 = 4
\ee

\be
\gamma_2 =  \frac{202}{3}-\frac{20}{9}N_f-\frac{20}{3} n_{\tilde{g}} 
\ee

\be
\gamma_3 = 1249-\biggl(\frac{2216}{27}+\frac{160}{3}\zeta(3)\biggr)N_f
                -\frac{140}{81}N_f^2
          -n_{\tilde{g}} \biggl(\frac{3566}{9} + \frac{280}{27}N_f\biggr)
                -\frac{140}{9}n_{\tilde{g}}^2 
\label{finalgammanum} 
\end{eqnarray}

We see that the gluino gives a substantial contribution to two- and 
especially three-
loop levels. Indeed, the gluino contribution reduces the two-loop coefficient by 
about 10\% and reduces the three-loop contribution by about 50\%. Such a large
contribution might ultimately be important in phenomenological applications.

The anomalous dimension of quark mass along with the QCD $\beta$-function
determines
the running of the quark mass. Indeed, at the three-loop level, for the running
quark mass one has (see, e.g., \cite{PL_H}) 
\begin{equation}
\frac{m_f(\mu_1)}{m_f(\mu_2)}=\frac{\phi(\alpha_s(\mu_1))}
                                   {\phi(\alpha_s(\mu_2))},
\label{eq:mrun}
\end{equation}
where,
\begin{eqnarray}
\lefteqn{\phi(\alpha_s(\mu))=\biggl(-2\beta_1\frac{\alpha_s(\mu)}{4\pi}\biggr)
                              ^{-\frac{\gamma_1}{\beta_1}}
       \biggl\{1
     -\biggl(\frac{\gamma_2}{\beta_1}
         -\frac{\beta_2\gamma_1}{\beta_1^2}\biggr)\frac{\alpha_s(\mu)}{4\pi}}
                                                             \nonumber\\
 && \quad \hspace{-3mm}
     +\frac{1}{2}\biggl[\biggl(\frac{\gamma_2}{\beta_1}
         -\frac{\beta_2\gamma_1}{\beta_1^2}\biggr)^2
       -\frac{\gamma_3}{\beta_1}
       +\frac{\beta_2\gamma_2}{\beta_1^2}+\frac{\beta_3\gamma_1}{\beta_1^2}
       -\frac{\beta_2^2\gamma_1}{\beta_1^3}
         \biggr]\biggl(\frac{\alpha_s(\mu)}{4\pi}\biggr)^2\biggr\}
\label{eq:f}
\end{eqnarray}
The above equation indicates that there will be a substantial shift in the
running of a quark mass due to the gluino.

To verify this, we will need the following equations.
The running coupling is parametrized as follows:
\begin{equation}
\frac{\alpha_s(\mu)}{4\pi}=-\frac{1}{\beta_1 L}-\frac{\beta_2 \log L}
{\beta_1^3 L^2}-\frac{1}{\beta_1^5 L^3}(\beta_2^2 \log^2 L-\beta_2^2 \log L
+\beta_3 \beta_1-\beta_{2}^{2})+O(L^{-4}),
\label{eq:Asparametr}
\end{equation}
where $L=\log(\mu^2/\Lambda_{\overline{MS}}^2)$.

The general evolution equation
for the running coupling to $O(\alpha_s^3)$ \cite{PL_H} has the form
\newpage
\begin{eqnarray}
\lefteqn{\hspace{-1cm}\frac{\alpha_s^{(n)}(\mu)}{4\pi}
                         =\frac{\alpha_s^{(N)}(M)}{4\pi}
      -\biggl(\frac{\alpha_s^{(N)}(M)}{4\pi}\biggr)^2
      \biggl(\beta_1^{(N)}\log\frac{M^2}{\mu^2}
      -\frac{2}{3}\sum_{l}\log\frac{m_l^2}{\mu^2} \biggr)}
                                                             \nonumber\\
 &&   -\biggl(\frac{\alpha_s^{(N)}(M)}{4\pi}\biggr)^3
      \biggl[\beta_2^{(N)}\log\frac{M^2}{\mu^2}
      -\frac{38}{3}\sum_{l}\log\frac{m_l^2}{\mu^2}
                                                             \nonumber\\
 && \quad \hspace{23mm}
      +\biggl(\beta_1^{(N)}\log\frac{M^2}{\mu^2}
      -\frac{2}{3}\sum_{l}\log\frac{m_l^2}{\mu^2} \biggr)^2
                           +\frac{50}{9}(N-n)\biggr]
\label{eq:Astransform}
\end{eqnarray}

where the superscript $n$ ($N$) indicates that the corresponding quantity
is evaluated for $n$ ($N$) numbers of
participating quark flavors.
Conventionally, $n$ ($N$) is specified to be
the number of quark flavors with mass $\leq \mu$ ($\leq M$). However, the
eq.(\ref{eq:Astransform}) is relevant for any $n\leq N$ and arbitrary
$\mu$ and $M$, regardless of the conventional specification of the number of
quark flavors. The $\log m_l/\mu$
terms are due to the ``quark threshold'' crossing effects and the constant
coefficients $2/3=\beta_1^{(k-1)}-\beta_1^{(k)}$,
$38/3=\beta_2^{(k-1)}-\beta_2^{(k)}$ represent
the contributions of the quark loop in the $\beta$-function.
The sum runs over $N-n$ quark flavors (e.g., $l=b$ if $n=4$ and $N=5$).
Note that $m_l$ is the pole mass of the quark with flavor $l$.
Quark masses can be estimated from QCD sum rules. For instance, 
the b quark pole mass is $m_b=4.72$GeV \cite{mb}.
  
In fig.2 we show the evolution of $m_b(\mu)$ from $\mu$=4.72GeV to $M_Z$.
We see that the gluino effect is a few percent at $M_Z$ which could 
ultimately be important for grand unification studies. If the gluino is not
light but is nevertheless below the squark mass, our current result will
still have a region of relevance. Ultimately, of course, it will be desirable
to have the full SUSY three-loop effect including squark contributions.

\begin{figure}
\hskip 3.0cm
\epsfxsize=3.5in \epsfysize=2in \epsfbox{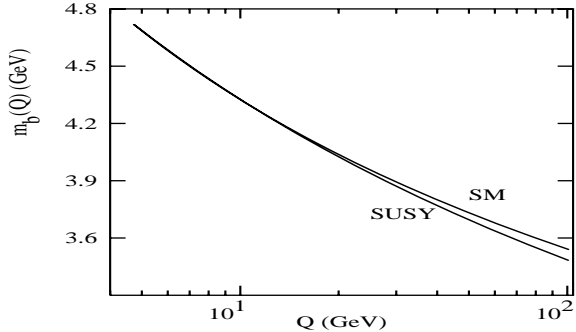}
\caption{The gluino effect on the running of the b quark mass}
\label{Fig.2}
\end{figure}

Analyzing numerical results, we see that the third-loop effect is only a few
MeV. Based on this, we conclude that the three-loop anomalous dimension
with the gluino included is a good approximation for phenomenological
applications and the error in the perturbative evaluation of this quantity
can be estimated at few MeV.
 
An observable quantity $R$ is invariant under the renormalization group
transformations and obeys the homogeneous renormalization group equation:
\begin{equation}
\biggl(\mu^2\frac{\partial}{\partial\mu^2}
   +\beta(\alpha_s)\alpha_s\frac{\partial}{\partial\alpha_s}
   -\gamma_m(\alpha_s)\sum_{f}
        m_f\frac{\partial}{\partial m_f} \biggr)
    R(\mu, \alpha_s, m_f)=0
\label{eq:RG}
\end{equation}
The quantity $R$ may denote various cross sections and decay rates that are
usually calculated using the perturbation theory method. In our previous works
we have calculated the gluino contribution to the QCD $\beta$-function at the 
three-
loop level \cite{betagluino} and to the hadronic decay rate of the Z boson 
to the four-loop level \cite{Zgluino}. Thus, the present work completes the
evaluation of gluino contributions necessary for $O(\alpha_s^3)$ 
renormalization group analysis for the above quantity. These results can be
used in the renormalization group analysis for other quantities, for instance,
hadronic decay rates of the $\tau$-lepton and various Higgs bosons.

\acknowledgements

We thank Phil Coulter for useful discussions. This work was supported
by the US Department of Energy under grant no. DE-FG02-96ER-40967.

\end{document}